# Sample NLPDE and NLODE Social-Media Modeling of Information Transmission for Infectious Diseases: Case Study Ebola

Armin Smailhodvic[1], Keith Andrew[1,4], Lance Hahn[2,4], Phillip C. Womble[1,4], Cathleen Webb[3]

[1]Departments of Physics and Astronomy, [2]Psychological Sciences, [3]Chemistry, and the [4]Cyber Defense Laboratory, Western Kentucky University
Bowling Green, KY 42104

We investigate the spreading of information through Twitter messaging related to the spread of Ebola in western Africa using epidemic based dynamic models. Diffusive spreading leads to NLPDE models and fixed point analysis yields systems of NLODE models. When tweets are mapped as connected nodes in a graph and are treated as a time sequenced Markov chain, TSMC, then by Kurtz's theorem these specific paths can be identified as being near solutions to systems of ordinary differential equations that in the large N limit retain many of the features of the original Tweet dynamics. Constraints on the model related to Tweet and re-Tweet rates lead to different versions of the system of equations. We use Ebola Twitter meme based data to investigate a modified four parameter model and apply the resulting fit to an accuracy metric for a set of Ebola memes. In principle the temporal and spatial evolution equations describing the propagation of the Twitter based memes can help ascertain and inform decision makers on the nature of the spreading and containment of an epidemic of this type.

## 1    Introduction

The expanding role of Social Media allows for unique opportunities to observe information transmission among digitally linked groups and populations[1]. For specific populations social networks can play a fundamental role in the spread of information, ideas, and trends[2] among its members[3]. The study of information transmission as a stochastic diffusion based process has been of interest to scientists for several years[4,5,6]. Such diffusion processes have been used to characterize collective behavior[7], adoption of innovations[8], spread of deviant behavior[9], trends in a city[10], social upheaval[11,12], computer virus propagation[13], etc. among a fixed population of users[14]. Similar models have also been developed independently in other domains[15] to study disease based epidemics[16,17], notably SARS[18,19] the H1N1[20,21] virus, and the synchronization of biological systems[22]. In the case of Twitter[23,24,25] and Instagram[26,27] there are ongoing studies about the speed of transmission, tracking[28], truthfulness, and the clustering of social memes[29]. For the case of disease transmission the benefits from epidemic modeling are three-fold: understanding possible mechanisms of spread of epidemics[30], predicting their future course in a population and developing strategies to effectively control them[31]. Recently, epidemic algorithms[32] have been proposed as a means to disseminate information in large scale settings, such as on the Internet, or in "Peer-to-Peer" networks. Such algorithms operate by letting desired information spread in a distributed system as an epidemic would spread throughout a group of susceptible individuals. The connected graph formed from the nodal graph of tweets and retweets for a particular meme gives rise to a chain of connected nodes with an information transmission structure that evolves in time. For a particular path each node is linked to the previous temporal node that can be approximated by a Markov chain when other nearest neighbors are ignored. In this case a particular path can be represented by a differential equation whose solutions are a model for information transfer amongst the nodes. We are especially interested in information transfer expressed as memes related to contagious diseases as given by Tweets in Twitter.

Ebola hemorrhagic fever is a highly infectious and lethal disease named after a river in the Democratic Republic of the Congo where it was first identified in 1976[33]. Twelve outbreaks of Ebola have been reported in Congo, Sudan, Gabon, and Uganda as of September 14, 2003. By September 14, 2014, a total of 4507 confirmed and probable cases of Ebola Virus Disease (EVD), as well as 2296 deaths from the virus, had been reported from five countries in West Africa — Guinea, Liberia, Nigeria, Senegal, and Sierra Leone. In terms of reported morbidity and mortality, the current epidemic of EVD is far larger than all previous epidemics combined. The true numbers of cases and deaths are certainly higher. There are numerous reports of symptomatic persons evading diagnosis and treatment, of laboratory diagnoses that have not been included in national databases, and of persons with suspected EVD who were buried without a diagnosis having been made[34].

The latest epidemic began in Guinea during December 2013, and the World Health Organization (WHO) was officially notified of the rapidly evolving EVD outbreak on March 23, 2014. On August 8, the WHO declared the epidemic to be a "public health emergency of international concern." By mid-September, nine months after the first case occurred, the numbers of reported cases and deaths were still growing from week to week despite multinational and multisectoral efforts to control the spread of infection. The epidemic has now become so large that the three most-affected countries — Guinea, Liberia, and Sierra Leone — face enormous challenges in implementing control measures at the scale required to stop transmission and to provide clinical care for all persons with EVD[35].

Because the Ebola virus is spread mainly through contact with the body fluids of symptomatic patients, transmission can be stopped by a combination of early diagnosis, contact tracing, patient isolation and care, infection control, and safe burial[36]. Before the current epidemic in West Africa, outbreaks of EVD in central Africa had been limited in size and geographic spread, typically affecting one to a few hundred persons, mostly in remote forested areas. The largest previous outbreak occurred in the districts of Gulu, Masindi, and Mbarara in Uganda. This outbreak, which generated 425 cases over the course of 3 months from October 2000 to January 2001, was controlled by rigorous application of interventions to minimize further transmission — delivered through the local health care system, with support from international partners[37].

By that time several mathematical models were proposed to analyze the spatial and temporal spreading of the disease. Population dynamics models with a healthy but susceptible population, S, focused on predicting the number of infected individuals, I, and recovered or removed individuals, R, due to vaccines or cures, these are often referred to as SIR based models. For instance Evans[38] and Lukcza[39] proposed a modified susceptible – infected – removed (SIR) disease model to capture the effect of the spreading individuals. The goal of this paper is to demonstrate that the spreading of social-media information for the West African Ebola epidemic can be of value in assisting to predict the spreading of infectious diseases.

In this paper, in Section 2 we review the basic models of immediate interest: the NLPDE Logistic Model, the SIS Model, and the SIR Model, in Section 3 we use a Meme Scoring indicator to evaluate Ebola memes and use the modified classical SIR disease model to capture the effects of Twitter based social-media diffusion and spreading, in section 4 we apply these models to the Ebola outbreak and conclude in Section 5.

## 2   Markov Processes and Approximate Differential Equations

The construction of a model based upon differential equations from data in a connected graph that is represented by a set of nodes has been quantified in detail in a Theorem by Kurtz and applied to epidemics and networks by Aspirot[40] and Venkatramanan[41]. In particular, for a Twitter based network composed of a set of time sequenced nodal events that can be represented as a Markov process[42], where each node depends only on the current node state and not on the prior history, then the path can be approximated by a solution to a differential equation as long as the large N limit of the discrete solution obeys the Lipshitz Condition, is Uniformly Convergent in the large N limit, has a Bounded Noise Variance and exhibits Convergence of the Initial Conditions, IC- stable initial conditions, IC, so that the large N limit gives ICs evaluated at the initial time that have the same information on the graph and for the continuous function; then in the limit that N→∞ the process can be approximated by a the solution to a system[43] of ODEs. Here we will ignore a detailed analysis of the topological effects in developing the system of equations[44] and overall network modularity[45] but the error bounds for the parameters are often difficult to determine with accuracy[46].

   A. **Finite- Bounded Diffusive PDE Logisitc S → I Model**
      We follow the model of Wang, Wang and Xu[47,48,49] based on a Fick's Law analysis to have spatial diffusion[50] and temporal population change[51] where we let I(x,t) be the number of infected individuals, I, and if the population is closed and finite so that the infected plus the susceptibles, S(x,t), obeys the constraint conservation equation: S(t)+I(t)=N, the distances *l* and L are the minimum and maximum distances from the initial source to the spatially[52] extended group of users, d represents the aggressiveness or diffusion constant of the spreading infection in our 1-d model space, the x coordinate represents the scaled spatial extent following the transmission of the disease,

the Neumann boundary conditions guarantees that there is no transmission beyond the edges of x, between *l* and L, then the minimum speed of propagation for reaction diffusion equations of this type is given by v, where the average node to node time is <Δt>, then the resulting solution reaches saturation of infected individuals in a finite time $T_c$, for an infection rate or contact rate in terms of number per unit time given by β then:

$$S(x,t) \to I(x,t) \quad N = S(t) + I(t) \quad \frac{\partial S}{\partial t} + \frac{\partial I}{\partial t} = 0$$

$$\text{Infection - Rate}: [\beta] = s^{-1} \quad \text{diffusion}: [d] = m^2 s^{-1}$$

$$\frac{\partial I}{\partial t} = d \frac{\partial^2 I}{\partial x^2} + \beta I(N - I) \quad I(x,1) = \phi(x)$$

$$l \leq x \leq L \quad \frac{\partial I(l,t)}{\partial x} = \frac{\partial I(L,t)}{\partial x} = 0 \quad t \geq 1$$

$$speed: v = \sqrt{2d \left(\frac{\partial f}{\partial t}\right)_{t=0}} \approx \sqrt{\frac{2d}{\langle \Delta t \rangle}} \approx \sqrt{2d\beta} \quad f = I(x,t)(N - I(x,t))$$

$$d = 0 \to \frac{dI}{dt} = \beta SI = \beta I(N - I) \to I(t) = \frac{I(0)N}{I(0) - (N - I(0))e^{-\beta Nt}}$$

$$T_C = \inf\{t : I(t) < N - 1\} \to T_C = \frac{1}{\beta N} \left\{\frac{(N-1)(N - I(0))}{I(0)}\right\}.$$

(1)

The diffusion constants for disease spread generally are on the order of 0.034 $m^2$/s giving ground based disease based average velocity for a homogeneous one dimensional population of over 1 km/day without boundary conditions for hospitals or border containments and it is difficult to translate this directly to a network topology. The temporal solution is a Sigmoid function indicating that in a finite critical time $T_c$ the population saturates, at a fixed time for large N we see that the spatial change is exponentially decreasing from the source displaying a decrease in the infection rate the farther away from the source for a noninfected individual. The spatial speed of propagation depends on how aggressive the disease is and the contact rate. The space independent, d=0, logistic equation, when converted into an iterative map, exhibits a rich structure associated with the nonlinearities inherent in the SI coupling term. There are parameter regions where the system exhibits a sensitive dependence on initial conditions and there are solutions that are chaotic[53].

B. **Kermack-McKendrick[54] S-I-S Model**

If the susceptibles and infectives have contact where the infectives can be cured or the disease has a finite lifetime then each population changes according to the Kermack-McKendrick system β is the contact rate and ν is the cure rate or 1/ν is the infectious period and the ratio β/ν is the basic dimensionless reproduction number:

$$S \to I \to S \quad N = S(t) + I(t) \quad \text{Infection-Rate}: \beta\, s^{-1} \quad \text{Cure-Rate}: v\, s^{-1}$$

$$\frac{dS}{dt} = -\frac{\beta}{N} SI + v I$$

$$\frac{dI}{dt} = \frac{\beta}{N} SI - v I \tag{2}$$

$$\text{If } \beta > v \quad \text{then} \quad \lim_{t \to \infty} S(t) = \frac{v N}{\beta} \quad \lim_{t \to \infty} I(t) = \frac{(\beta - v)N}{\beta}$$

$$\text{If } \beta \leq v \quad \lim_{t \to 0} I(t) = 0 \quad R_{crit} = \frac{\beta}{v} \quad T_{infect} = \frac{1}{v}$$

here we consider the discrete trajectory $f_N = (S_N, I_N)$ as the path that in the limit becomes the continuous parametric function $f=(S(t),I(t))$ representing a deterministic epidemic model. As long as $R_o \leq 1$ the disease will be eradicated.

### C. Bounded SIR Model

The classical SIR disease[55] model splits a fixed-size population, N, into three distinct classes: the susceptible individuals, S, those do not have the disease and become infected; infected individuals, I, those who have the disease and can infect susceptible individuals; and the removed individuals, R, those who have recovered, die, or moved into isolation. Individuals in the removed class gain permanent immunity and remain in the R class forever.

Individuals in the population are assumed to homogeneously mix. Contacts between the susceptible and infected individuals result in a new infected individual at a rate proportional to the number of susceptible and infected individuals. Infected individuals are removed to class R at a rate proportional to the number of infected individuals. The system of nonlinear ordinary differential equations describing the SIR model is:

$$S \to I \to R \quad \text{Infection-Rate}: \beta\, s^{-1} \quad \text{Cure-Rate}: v\, s^{-1}$$

$$\frac{dS}{dt} = -\frac{\beta}{N} SI \quad S(0) = S_o < N$$

$$\frac{dI}{dt} = \frac{\beta}{N} SI - v I \quad I(0) = I_o < N - S_o \tag{3}$$

$$\frac{dR}{dt} = v I \quad N = S(t) + I(t) + R(t)$$

$$\lim_{t \to \infty} I(t) \to 0 \quad S(\infty) = \lim_{t \to \infty}(N - R(t)) = N - R(\infty) = S(0) e^{-\frac{\beta}{v} R(\infty)}$$

$$R_{crit} = \frac{\beta S(0)}{v N} > 1 \to epidemic$$

where β is the transmission rate and ν is the cure or removal rate. In the long term limit of this model there are no remaining infected individuals. There exists a critical value for the removal rate required for epidemic spread to occur, if the initial size of the susceptible population is too small then an epidemic will not break out. Although these equations can be modified for the natural birth and death rates for the population of interest these time scales are too long to be considered for our analysis.

## 3    The Twitter based STR model for Markov Memes

To track memes of interest we used the meme score method to capture memes of interest for Ebola tracking on Twitter following the procedure established by Kuhn, Perc and Helbing[56]. The initial meme concept arises from an idea of Dawkins[57] in his development of language word constructions that have cultural and social features similar to genes[58] and embody word change[59]. Memes transfer information, cluster[60] and can enter into conventional usage on short time scales. To more accurately capture the nature of a given meme we use the n-gram definition where an n-gram is a meme consisting of n related words. N-grams can be characterized by their relative entropy[61] and nearby word associations[62], some word combinations are much more probable and often certain pairings can be attributed to geolocation[63,64]. Here we consider our meme volume N to be the total number of tweets associated with our root, here it is Ebola, then $N_m$ is the number of occirances of the meme and $f_m$ is the frequency of occurrence of the meme of interest. We also use Kuhn's propagation weight $P_m$ which measures whether or not the community of users found it interesting. Then the number of Twitter accounts that received the n-gram and then retweet the n-gram at least once is $n_{rtw}$ and the total number of tweets with the n-gram is $n_{tw}$, likewise the number of accounts where the tweet stays and it is not retweeted is $s_{nrtw}$ while the total number that retweet or do not retweet is $s_{tnrtw}$ thereby giving the meme score as:

$$M_m = f_m P_m = \frac{N_m}{N}\left(\frac{n_{rtw}/n_{tw}}{s_{nrtw}/s_{tnrtw}}\right) \qquad . \quad (4)$$

We distinguished between tweets from different countries with a focus on Sierra-Leone, Liberia, and Guinea.

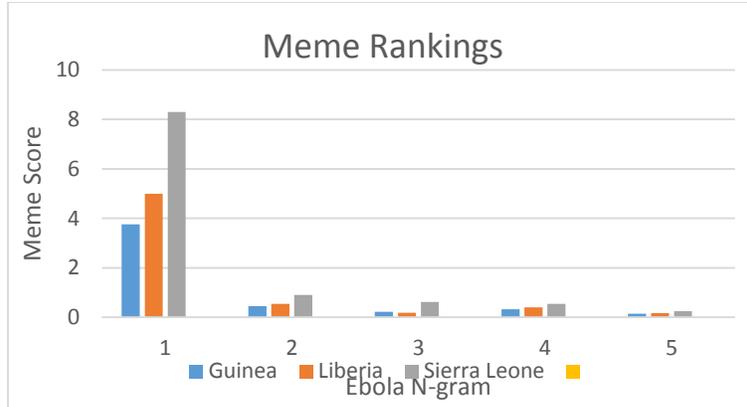

Figure 1. Meme Scores for n-grams with the 1-gram meme Ebola as the lead term corresponding to column 1, other are n-grams related to sickness, death, illness, and help in separate N-grams, the Scores are scaled by a factor of $10^4$.

Following Venkatramanan and Kumar we consider the nodal Tweet graph as a Discrete Time Markov Chain that obeys the Kurtz Theorem. Then Venkatramanan proves that the Homogeneous Influence Linear Threshold (HILT) model[65] for information transmission in the continuous fluid limit for a network is equivalent to a version the SIR epidemic disease spread model. Following Jin, Dougherty, Saraf, Cao and Ramakrishnan[66] we first adopt a special modified SIR model and then switch to a tweet focused model to track Ebola Tweet information transmission. One advantage of the expanded model is that it allows for the development of a potential metric as a measure to assist in distinguishing between the propagation of a rumor[67] vs. a true news story which we apply to Ebola information. First we adopt the Twitter based Susceptible-Tweet-Retweet[68], STR model, a modification of the classical SIR model, which captures some aspects of social-media information spreading. In the STR model we compare 1-gram base memes to either 2-gram, 3-gram, or 4-gram memes. The idea of using n-gram memes for word comparisons has been used extensively by Hahn et. al. In the STR model, S is the total number of individuals capable of receiving a Tweet related to the base 1-gram meme which we take initially to be our population. Then T, is the total number of 2-gram or 3-gram social-media Tweeted individuals, and R represents the total number of individuals who retweet any version of the meme. Twitter diffusion rates depend heavily on the meme socre more so than on the topic meme rating[69]. As with the SIR model this means that for a finite population, in the long time limit, R will saturate and be a Sigmoid function and that the n-gram information number of individuals, T, will peak and eventually trend to zero as a function of time. This model can be described with the following system of differential equations:

$$S \to T \to R \quad Tweet-Rate: \beta\ s^{-1} \quad Retweet-rate: \nu\ s^{-1}$$

$$\frac{dS}{dt} = -\frac{\beta}{N} ST$$

$$\frac{dT}{dt} = \frac{\beta}{N} ST - \nu T \quad (5)$$

$$\frac{dR}{dt} = \nu T \quad N = S(t) + T(t) + R(t)$$

$$\lim_{t \to \infty} T(t) \to 0 \quad \lim_{t \to \infty}(N - R(t)) = N - R(\infty) = S(0) e^{-\frac{\beta}{\nu} R(\infty)}$$

where the total number of users remains a fixed population, N, and the number of primary Tweeter users T climbs, reaches a peak and then asymptotically goes to zero, as S decreases and stabilizes and R grows and stabilizes to asymptotic values that add up to N.

**Four Variable Model and Metric**

A model that includes delayed response, nonrepsonders and has been used to analyze rumors[66,70] and news information spreading on social networks and adds another population. Based on this reasoning, we considered a more applicable, the SEIZ model which was first used to study the adoption of Feynman diagrams and used by Venkatramanan[71] is the one we use here. In the context of Twitter, the different compartments of the STRZ model are identified as: Susceptible (S) represents a user who has not received any Tweet; Tweet (T) denotes a user who has tweeted about the event; Zero-Tweets (Z) is a user who has heard about the news but chooses not retweet about it; and Re-Tweet (R) is a user who has received the news via a tweet and retweets later. This terminology is widely used and very similar to the original authors[72] of the SEIZ model.

$$S \to E \to I \to Z$$

$$\frac{dS}{dt} = -\frac{\beta}{N} SI - \frac{b}{N} SZ$$

$$\frac{dT}{dt} = p \frac{\beta}{N} ST + \left(\frac{\rho I - N\varepsilon}{N}\right) R$$

$$\frac{dR}{dt} = (1-p) \frac{\beta}{N} ST - (1-l) \frac{b}{N} SZ - \left(\frac{\rho I - N\varepsilon}{N}\right) R \quad (6)$$

$$\frac{dZ}{dt} = \frac{lb}{N} SZ$$

$$Robustness-Metric: R_{Rob} = \frac{(1-p)\beta + (1-l)b}{\rho + \varepsilon}$$

| | |
|---|---|
| $\beta$ | S-T contact rate |
| b | S-Z contact rate |
| $\rho$ | R-T contact rate |
| $\varepsilon$ | Incubation rate |
| $1/\varepsilon$ | Average Incubation Time |
| bl | Effective rate of S → Z |
| $\beta\rho$ | Effective rate S → T |
| b(1-l) | Effective rate S → R via contact with Z |
| $\beta$(1-p) | Effective rate S → R via contact with T |
| l | Probability S → Z given contact with skeptics |
| (1-l) | Probability S → R given contact with skeptics |
| p | Probability S → T given contact with adopters |
| (1-p) | Probabiity S → R given contact with adopters |

We also use the robustness ratio[66] $R_{Rob}$, as a ratio between the total incoming and outgoing tweets, the contributions from the susceptible and tweeted compartments in this quantity. A $R_{Rob}$ value greater than 1 implies that the influx into the retweeted compartment is greater than the efflux. Similarly, a value less than 1 indicates that members are added to the retweeted group more slowly than they are removed. This metric was useful in determining whether or not an event was news or a rumor, in general it remained high for news events and was low for rumor events.

## 4      A Data Application to Western Africa Ebola

With the Ebola virus the incubation period is between 2 and 21 days, this would mean that infected individuals could travel with no symptoms for this time and they are not infectious until the incubation is over. Twitter parameters are difficult to determine depending on the region one is using for their Twitter mining. In countries where the Internet is widely available, information is shared within a day. In countries where the Internet is sparser, information sharing can take up to three days. Thus the Tweet incubation time value can range between 1 and 0.33 days. It is not possible to determine how many of the people with Twitter accounts in the regions of interest will be part of the Ebola discussion so the exact initial numbers are treated as parameters that fit a curve to the data. In terms of Ebola Tweets in the three countries for a month we find several million, but only about 10% deal with the memes of interest.

Table 1. Tweets for the month of August 2014

| Ebola Tweets (S) | Ebola Liberia Tweets (T) | Ebola Sierra Leone Tweets (T) | Ebola Guinea Tweets (T) |
|---|---|---|---|
| 6,582,123 | 407,809 | 152,867 | 61,371 |

To convert the data to useable parameters for the STR Mathematica model, the variables normalized to a total population size of 1.0 and the STR variables are densities.

Table2. Normalized parameters.

| S | Liberia (T) | Sierra Leone (T) | Guinea (T) |
|---|---|---|---|
| 1 | 0.062 | 0.023 | 0.0093 |

The Mathematica input parameters are adjusted to meet the data being interpreted, numerical solutions were generated using the adaptive NDSolve routine with interpolating functions and out as parametric plots. Initial values and the parameters β, ν and d were varied to get the best fit to the data from each country over a 30 day period and then were used to match the next period. The STR Twitter model was then compared to the observable Ebola cases in order to compare the correlation of Twitter data to the spread of the disease. The following graphs depict the individual linear equations along with the standard error associated with each new case calculation.

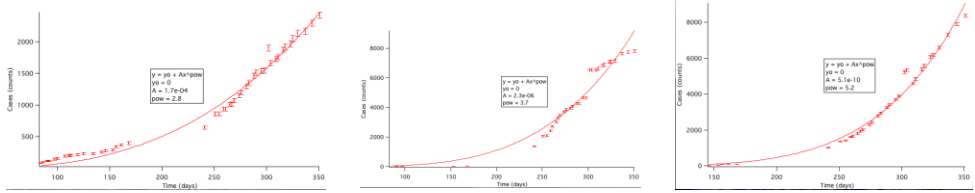

**Figure 1 STR vs. time**

Figure 2. Data match to determine optimal values of beta for Sierra Leone, Guinea and Liberia plotting the number of Tweets vs 350 days of observations starting with a power law fit.

The same process was then repeated for the twitter data mined for Sierra Leone and Guinea.

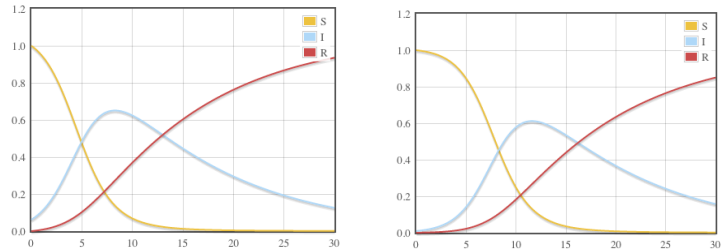

Figure 3. Normalized, N=1, NDSolve solutions for Ebola Liberia showing S, T and R values vs. day for 30 days normalized to N=1 showing characteristic Sigmoid curves with T exhibiting the decay to zero over long times.

Based on these calculations we have been able to predict new cases within a 10% range of the new cases documented by the World Health Organization and the Centers for Disease Control.

Each model provides its own insight into the meme spreading: the Logistic S→I Model gives a diffusion speed and a finite critical saturation time, the S → I → S Model gives a direct reproduction factor R, the STR model also yields a value for R and the STRZ model has a robustness parameter that is an indicator that helps to distinguish rumors from news in many settings. Using these parameters in the renormalized population models for each country we find approximate values of for the robustness metric.

Table 3. Parameters for the different models.

|  | β | STRZ R | STRZ R robust |
|---|---|---|---|
| **Guinea** | 0.00493 | 1.072 | 12.73 |
| **Liberia** | 0.00714 | 1.284 | 9.51 |
| **Sierra Leone** | 0.00302 | 1.654 | 6.22 |

All of our values of the dimensionless reproduction factor R remain greater the one. In general larger robustness values correspond to information similar to news coverage while low numbers correspond to rumor levels of information, i.e. < 1.0. These values give some insight into how these models provide interesting and useful information for the special case of using Twitter to follow Ebola in Western Africa. Also of interest is how the reproduction number R, has been changing with time. In some sense this is a strong indicator of whether or not the epidemic is coming under control. If we select different months and finite differencing to average over then the rate of change can be found:

Table 4. Rates of Change of the Reproduction factors.

| $x10^{-3}$ | Guinea | Liberia | Sierra-Leone |
|---|---|---|---|
| $dR/dt$  $d^{-1}$ | 1.2 | 4.2 | 17.6 |
| $d^2R/dt^2$  $d^{-2}$ | 0.01 | 0.2 | 2.6 |

Figure 4. Numerical estimates of the first and second derivatives of the overall production rate taken over successive months in a 120 day window.

Currently in each country R>1 indicating ongoing epidemics. When $dR/dt < 0$ the spreading will be dropping, as such the rate of change of dR/dt is also an indicator of how the disease is coming under control. One way to introduce this change is to have the coupling constants have an explicit time dependence in the models.

## 5     Conclusions

These results were used in the continuous prediction process of the spreading of the Ebola virus in West Africa. The predictions were used in order to determine the spread of the virus into surrounding countries as well as predictions on the new amount of cases each week. These predictions were then compared to the values posted by the Centers for Disease Control and Prevention and the World Health Organization. The predictions during the first couple of months for the spread of the Ebola virus were within 10% of the accepted value, which was the value given by the Centers for Disease Control and Prevention and the World Health Organization.

Since the Ebola virus has crossed continental borders the background noise has increased exponentially and in turn has created difficulties in the data mining process. As an example before the virus was feared of spreading throughout the United States, the average amount of data mined was 6.5 million tweets; but after fear of the virus spreading throughout the United States the average amount of data mined was 20.3 million tweets. The reduction background noise over time will continue to improve the data set and in turn increase the accuracy of the predictions.

The one single factor that would increase the accuracy of the case predictions and possible tracking of any virus would be for all twitter users to enable GPS location services, since currently only 3% of twitter users have GPS location services enabled a plethora of data is not being used for analysis. By enabling GPS services more data could be collected allowing for more accurate tracking and possible prevention of any virus spreading throughout a population. Current software allows for 78% accuracy on predicting where a tweet originated from, this is based on the country. For example if one were to try and data mine tweets in the United States based on states that 78% accuracy would drop to 24%.

New efforts are underway to understand the nature of the Ebola spread and resiliency. Models are being developed to examine community interactions, the effects of several strains, time dependent couplings, genuine wave equation interactions, etc. Many of these efforts are based upon first person accounts collected by workers in the region who had firsthand knowledge and exposure to the community during the efforts to institute treatments. The nature of the Ebola carriers has also complicated matters with containment where it appears that several species of bats, especially the common fruit bat, have a genome that makes them excellent carriers. Also of great service is the development of a vaccine now in clinical trials that will impact the spread of the disease and the nature of modeling the epidemic across the world. The social media archives will provide a rich data set to assist in the reconstruction of our understanding of the spread of the disease and wills serve as a permanent and accessible aid in our understanding of the transmission of information.

Acknowledgements: We thank the WKU Cyber Defense Laboratory and Electronic Warfare Associates for generous support and access to unique code and search methodologies (ICGN: 2178539-57-cyhv.3).6       References

[1] H. Kwak, C. Lee, H. Park, and S. Moon. What is twitter, a social network or a news media? In Proceedings of the 19th International Conference on World Wide Web, pages 591–600. ACM, 2010
[2] Sheth, Amit, Hermant Purohit, Ashutosh Jadhav, Pavan Kapanipathi, and Lu Chen. "Understanding events through analysis of social media." *Proc. WWW 2011* (2010).
[3] E. Bakshy, J. Hofman, W. Mason, and D. Watts. Everyone's an influencer: quantifying influence on twitter. In Proceedings of the 4th ACM International Conference on Web Search and Data Mining, pages 65–74. ACM, 2011
[4] M. Cataldi, L. Di Caro, and C. Schifanella. Emerging topic detection on twitter based on temporal and social terms evaluation. In Proceedings of the 10th International Workshop on Multimedia Data Mining, page 4. ACM, 2010
[5] Agreste, Santa, Pasquale De Meo, Emilio Ferrara, Sebastiano Piccolo, and Alessandro Provetti. "Analysis of a heterogeneous social network of humans and cultural objects." *arXiv preprint, arXiv:1402.1778* (2014).
[6] Thelwall, Mike, Stefanie Haustein, Vincent Larivière, and Cassidy R. Sugimoto. "Do altmetrics work? Twitter and ten other social web services." *PloS one* 8, no. 5 (2013): e64841.12